\documentclass[twocolumn,prl,aps,superscriptaddress,floatfix,showpacs]{revtex4}

\usepackage{bm}
\usepackage{epsfig}
\usepackage{graphicx}
\DeclareGraphicsExtensions{.eps}
\usepackage{color}
\def\urs{URu$_2$Si$_2$}

\begin{document}

\title{Similarity of the Fermi Surface in the Hidden Order State and in the Antiferromagnetic State of \urs}

\author{E. Hassinger}
\email{elena.hassinger@gmail.com}
\affiliation{INAC, SPSMS, CEA Grenoble, 38054 Grenoble, France}

\author{G. Knebel}
\email{georg.knebel@cea.fr}
\affiliation{INAC, SPSMS, CEA Grenoble, 38054 Grenoble, France}

\author{T. D. Matsuda}
\affiliation{INAC, SPSMS, CEA Grenoble, 38054 Grenoble, France}
\affiliation{ASRC, Japan Atomic Energy Agency, Tokai, Naka, Ibaraki 319-1195, Japan}

\author{D. Aoki}
\affiliation{INAC, SPSMS, CEA Grenoble, 38054 Grenoble, France}

\author{V. Taufour}
\affiliation{INAC, SPSMS, CEA Grenoble, 38054 Grenoble, France}

\author{J. Flouquet}
\affiliation{INAC, SPSMS, CEA Grenoble, 38054 Grenoble, France}

\date{\today}

\begin{abstract}
Shubnikov-de Haas measurements of high quality \urs\ single crystals reveal two previously unobserved Fermi surface branches in the so-called hidden order phase. 
Therefore about 55\,\% of the enhanced mass is now detected. Under pressure in the antiferromagnetic state, the Shubnikov-de Haas frequencies for magnetic fields applied along the crystalline $c$ axis show little change compared with the zero pressure data. This implies a similar Fermi surface in both the hidden order and antiferromagnetic states, which strongly suggests that the lattice doubling in the antiferromagnetic phase due to the ordering vector Q$_{\mathrm{AF}} = $\,(0~0~1) already occurs in the hidden order. These measurements provide a good test for existing or future theories of the hidden order parameter.

\end{abstract}


\maketitle
The electronic properties of uranium compounds are determined by the tenuous balance between the localized and itinerant character of the $5f$ electrons which may lead to the formation of enigmatic ground states \cite{Flouquet2005}. One famous example is the heavy fermion compound \urs\ which shows a second order phase transition to a ``hidden order'' (HO) state at $T_0 = 17.5$\,K. The transition to the HO state is associated with a huge entropy loss of $0.2R\ln{2}$ \cite{Palstra1985}. Despite intense research for 25 years, the order parameter has not yet been identified. The possible proximity to a 5$f^2$ configuration of the uranium atoms leads to the possibility of multipolar ordering which is highly debated in Pr$^{3+}$ systems in the $4f^2$ configuration \cite{Sato2009}. Thus the resolution of the HO parameter will have a deep impact on the understanding of heavy fermion materials. A large diversity of theoretical proposals have been given. The most recent ones include multipolar orders \cite{Haule2009,Harima2010a,Cricchio2009}), dynamical spin density wave \cite{Elgazzar2009} or hybridization wave \cite{Dubi2010}.

The Fermi surface (FS) properties are directly linked to the itineracy of the 5$f$ electrons and to the change of the symmetry entering into the HO phase.
Changes of the FS at $T_0$ have been observed in various experiments. Optical conductivity \cite{Bonn1988} and transport measurements \cite{Schoenes1987,Maple1986} indicate a gap opening and a drop in the number of charge carriers at $T_0$. Recent STM measurements show that a hybridization gap opens suddenly at $T_0$ \cite{Schmidt2010,Aynajian2010} while in ARPES measurements abrupt changes of the electronic spectrum are detected \cite{Santander2009, Yoshida2010}.
Here we focus on the FS determination via Shubnikov-de Haas (SdH) measurements on a new generation of high quality crystals. SdH measurements under pressure provide the great opportunity to study the difference of quantum oscillations between the low pressure HO phase and the high pressure antiferromagnetic (AF) phase with propagation vector Q$_{\mathrm{AF}} = $\,(0~0~1) and ordered moment $m_0 = 0.3$\, $\mu_{\mathrm{B}}/$U. A small pressure of $P_x \approx 0.8$\,GPa is enough to switch the ground state from HO to AF \cite{Amitsuka2007,Hassinger2008,Butch2010}. The AF phase has been well characterized, notably the change from body centered tetragonal to simple tetragonal crystal structure below $T_0$ \cite{Elgazzar2009,Amitsuka1999}. Inelastic neutron scattering experiments under pressure suggest that, due to the disappearance of the inelastic signal at Q$_{\mathrm{AF}}$ and the emergence of an elastic signal at the same wavevector, the HO wavevector may also be Q$_{\mathrm{AF}}$ \cite{Villaume2008}. A test of this proposal is the persistence of the same FS through $P_x$. As proposed by recent band structure calculations in the AF state, the FS should show band folding \cite{Oppeneer2010}. Previous SdH measurements under pressure showed evidence of a similar FS on both sides of $P_x$ for only one frequency \cite{Nakashima2003}. Furthermore in the HO phase no proof of band folding has been provided up to now. The recent progress in the quality of the crystals allows us to present a detailed SdH study at ambient pressure as well as under high pressure. We show that the band folding appears already in the HO phase and present the pressure evolution of the different frequencies. 

High quality single crystals were grown using the Czochralski method in a tetra-arc furnace \cite{Aoki2010}. 
Sample 1 used for pressure measurements had an RRR = $\rho(300\mathrm{K})/\rho(2\mathrm{K})$ = 160 and sample 2 for ambient pressure measurements had an RRR = 175.
Standard four point ac-resistivity ($\rho$) measurements with a current of $I \approx 10$\,$\mu$A along the crystalline $a$ axis have been carried out in a dilution refrigerator with a base temperature of 20\,mK and in magnetic fields up to 13.2\,T. All the magnetoresistance curves and SdH spectra shown in this letter are taken below 35\,mK. The signal was amplified by a low temperature transformer by a factor of 1000 keeping the noise level very low.
Pressure up to 1.55\,GPa was applied with a NiCrAl-CuBe hybrid piston cylinder pressure cell with Daphne oil 7373 as the pressure transmitting medium. 
At $P=0$ the rotation angle was determined by the well established superconducting upper critical field $H_{\mathrm{c2}}$ curve \cite{Brison1995,Ohkuni1999}. 
$H_{\mathrm{c2}}$ is defined where $\rho = 0$. All angles are given as the deviation from the $c$ axis in the $ac$ plane. The field range of the normal state becomes smaller and consequently the resolution of the fast Fourier transform (FFT) is lowered when approaching $H \parallel a$. As the maximum field of the magnet was 13.2\,T, the limit for these measurements was around 80$^\circ$. Under pressure, measurements were performed only for $H\parallel c$. 

\begin{figure}
\includegraphics[width=65mm]{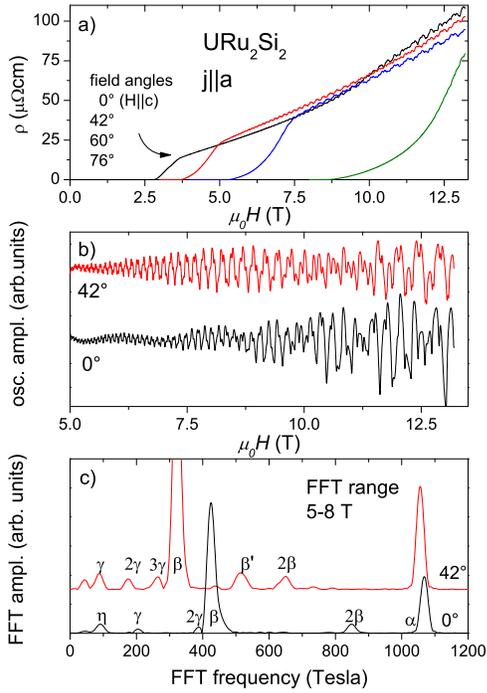}
\caption{(Color online) a) Magnetoresistance for several angles. b) SdH oscillations in resistivity when the polynomial background is subtracted for two angles. c) FFT spectra obtained for a field region of 5 - 8\,T for the same angles.}
\label{angdeprho}
\end{figure}

From the normal state resistivity (see Fig.\,\ref{angdeprho}a), a polynomial background is subtracted which leaves the SdH oscillations only (Fig.\,\ref{angdeprho}b). Performing a FFT we obtain spectra as shown in Fig.\,\ref{angdeprho}c. 
The cyclotron masses $m^\star$ were determined from the temperature dependence of the peak amplitudes which follow the Lifshitz-Kosevich formula as long as the peaks are well separated. The masses were determined in a field range 8\,-\,13.2\,T in order to have the same range for small and large angles and different pressures.
\begin{figure}
\includegraphics[width=87mm]{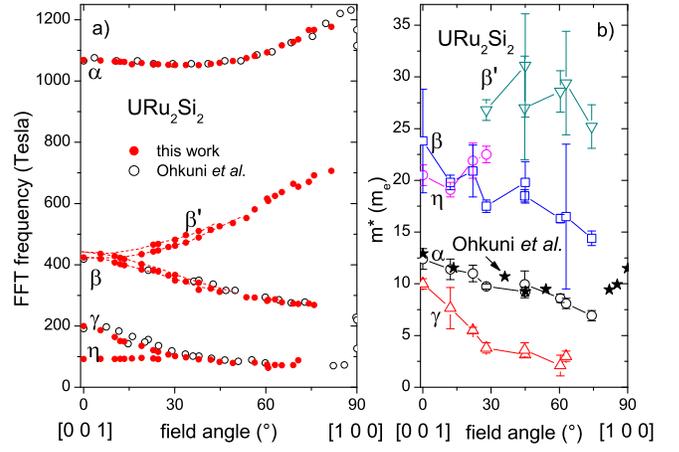}
\caption{(Color online) a) Angular dependence of the FFT frequencies obtained from SdH measurements in \urs\ reflecting the cross sectional areas of the different Fermi surface branches perpendicular to the magnetic field. The field range is 5\,-\,8\,T as long as $H_{c2}$ is low enough, then adapted to the available field range. b) Angular dependence of the effective masses. Earlier results by Ohkuni et al. \cite{Ohkuni1999} are also plotted and are in excellent agreement.}
\label{angdepfreq}
\end{figure}

Fig.\,\ref{angdeprho}c shows the FFT spectrum for $H \parallel c = 0^{\circ}$ in the field range 5 - 8\,T. In agreement with Ohkuni {\it et al.} \cite{Ohkuni1999} we found the $\alpha$, $\beta$ and $\gamma$ branches with frequencies $F_{\alpha} = 1065$\,T, $F_{\beta} = 425$\,T, $F_{\gamma} = 200$\,T and corresponding cyclotron masses $m^{\star}_{\alpha} = 12.4 m_e$, $m^{\star}_{\beta} = 23.8 m_e$, $m^{\star}_{\gamma} = 10 m_e$. However, we could also detect a band $\eta$ with small frequency $F_{\eta} = 93$\,T and $m^{\star}_{\eta} = 20.5 m_e$. The signal in the spectrum for even lower frequencies comes partly from the imperfect subtraction of the background. For $H \parallel c$ we see up to 9 harmonics for the $\alpha$ branch, four harmonics for $\beta$ and three for $\gamma$ and various combinations of different branches due to quantum interference.

The angular dependence of the oscillation frequencies is shown in Fig.\,\ref{angdepfreq}a. Only the fundamental branches are plotted. Contrary to previously published results \cite{Ohkuni1999} (black empty dots) additional branches are observed.
Most importantly, the $\beta$ branch splits into two branches $\beta$ and $\beta$' when rotation from $H \parallel c$ to $H \parallel a$. The previously unobserved branch $\beta$' has been observed in three different samples. 

The splitting into two branches means that the corresponding Fermi surface has different pockets with the same extremal cross sectional area for $H \parallel c$ and different areas for $H \parallel a$. 
For symmetry reasons it corresponds to a Fermi surface with four non-central flattened pockets along the main axes of the Brillouin zone (BZ). Such a Fermi surface appears in band structure calculations for a small moment AF phase \cite{Yamagami2000}(an obsolete idea for the HO phase), for the pressure induced large moment AF phase \cite{Elgazzar2009} but also in calculations in space group $P4_2/mmm$ (No. 136) \cite{Harima2010} in the paramagnetic state.
In these calculations, the flattened pockets are a product of the overlap of two large pockets at the $\Gamma$ and $Z$ point of the body centered tetragonal BZ, which are folded on top of each other in the simple tetragonal BZ, when the unit cell doubles with ordering vector Q$_{\mathrm{AF}}$. Therefore these pockets would not exist without ordering with Q$_{\mathrm{AF}}$. The fact that the folded Fermi surface in the AF state matches the Fermi surface in the HO state is a strong indication that the folding also takes place is the HO phase. 

Note that each of the $\beta$ and $\beta$' branches is split itself into two frequencies. Possible origins of this effect are that the spin up and spin down branches are split non-linearly with magnetic field or a warping of the corresponding Fermi surface pocket. Because the frequencies lie very close, two separated peaks appear only for some angles in the FFT. But the beating in the raw data (see Fig.\,\ref{angdeprho}b) is clear evidence, that even the peak for $H \parallel c$ must inhibit a second peak with small amplitude and very close frequency. The approximate position of this peak is schematically plotted as dashed line in Fig.\,\ref{angdepfreq}a. For higher angles and increasing $H_{\mathrm{c2}}$, the resolution of the FFT becomes too bad to decide whether two frequencies exist but the decreased beat frequency at 42$^\circ$ implies that the splitting must have decreased.
With increasing field, additional "side peaks" with unclear origin appear just above both the $\beta$ and $\beta$' branches. The detailed field dependence of the SdH spectra is a subject for further studies. This is the reason why only spectra for the lower field range $5$\,T$<H<8$\,T are shown here.

The frequency of the new heavy band $\eta$ is independent of the angle. It could only be observed for small angles. For higher angles above 30$^\circ$, where $F_{\gamma}$ is close to $F_{\eta}$ only one peak with a light mass in agreement with the $\gamma$ branch is observed.
In Fig.\,\ref{angdepfreq}b we show the angular dependence of the cyclotron masses. All masses besides $m^{\star}_{\eta}$ decrease with increasing angle. As we are not able to measure close to $H \parallel a$, we cannot exclude that the masses rise strongly in this direction due to the interaction of the electrons with the Ising-like longitudinal excitations. An increase for $H \parallel a$ has been observed for the $\alpha$ branch (black stars in Fig.\,\ref{angdepfreq}b) \cite{Ohkuni1999}. From the angular dependence of $H_{c2}$ it is expected to see higher masses for $H \parallel a$ than for $H \parallel c$.
Assuming spherical isotropic Fermi surfaces with extremal cross sectional area $S_{\mathrm{F}} = \frac{2\pi e}{\hbar} F=\pi k_{\mathrm{F}}^2$ where $F$ is the oscillation frequency, we can estimate the Sommerfeld coefficient $\gamma$ with the determined cyclotron masses $\gamma \approx \sum_i\frac{k_{\mathrm{B}}^2  V  m_i^*  k_{\mathrm{F}i}}{3\hbar^2}$, where $V = 49$\,cm$^3$/mol is the molar volume of \urs. Counting the heaviest branch $\beta$ four times and the other bands once, we obtain $\gamma \approx 37.5$\,mJ/molK$^2$. From specific heat a Sommerfeld term of $\gamma \approx 65$\,mJ/molK$^2$ has been measured \cite{Maple1986}. We have not detected any other branch. Such a branch should exist. Assuming that the mean free path is not much lower than in detected branches, the effective mass should be $m^{\star} > 70 m_e$ due to the sensitivity of our experiment. The modest $\gamma$ term and the low charge carrier concentration limit the size of this branch to the same order of magnitude as the already detected branches. It is predicted in Ref.~\onlinecite{Elgazzar2009} and required for compensation.

\begin{figure}
\includegraphics[width=87mm]{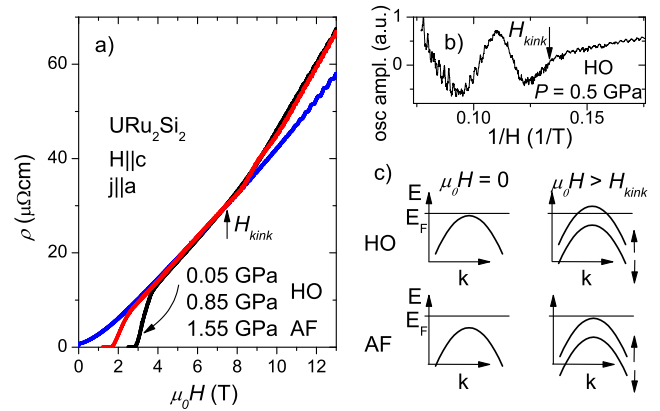}
\caption{(Color online) a) Pressure dependence of the magnetoresistance of \urs\ for three pressures in the whole measured field range. b) Low frequency oscillation appearing above the kink field $H_{kink}$. c) Schematic field dependence of a band in the HO and AF phases.}
\label{pdepmagnres}
\end{figure}
By applying pressure, the system switches from the HO phase to the AF phase. The Fermi surface properties in both phases were detected. Fig.\,\ref{pdepmagnres}a shows the magnetoresistance of \urs\ for different pressures. Up to $P=0.85$\,GPa, we see a clear kink at around $H_{kink} \approx 8$\,T. The kink was also observed in the two high quality samples we measured at ambient pressure in the same geometry (see Fig.\,\ref{angdeprho}a). The kink is smeared out very quickly with temperature and disappears above roughly 200 mK. Above $H_{kink}$, an oscillation with very low frequency $ f < 10$\,T appears (see Fig.\,\ref{pdepmagnres}b). In thermoelectric power a minimum appears at approximately the same field \cite{Malone2010}. It is an indication of a reordering of the Fermi surface possibly due to the polarization of a band for $H>H_{kink}$ as schematically presented in the upper part of Fig.\,\ref{pdepmagnres}c and thus entering into the framework of a Lifshitz transition.
The kink disappears for the last two pressures in the AF phase. This could result from the higher charge gap and therefore lower lying band in the AF phase as in the lower part of Fig.\,\ref{pdepmagnres}c. Additionally, the decreased mass under pressure (see below) leads to a decreased magnetic susceptibility and a decreased polarization with magnetic field.
In this sample, in the present pressure conditions, the critical pressure $P_x$ is near 0.85\,GPa. In relation to the disappearance of the kink, there is an abrupt change of the temperature dependence of resistivity at zero field and superconducting parameters $T_{\mathrm{SC}}$ and $H_{\mathrm{c2}}$ between 0.85\,GPa and 1.32\,GPa \cite{Hassinger2010b}. 

A previous pressure study of quantum oscillations could only follow the $\alpha$ branch with light mass \cite{Nakashima2003}. However, in the pressure experiment presented here, all the branches besides the $\eta$ branch could be detected. 
\begin{figure}
\includegraphics[width=83mm]{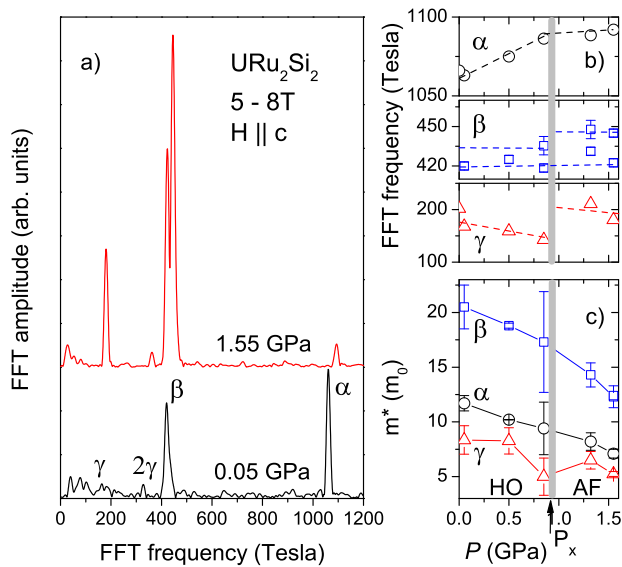}
\caption{(Color online) a) FFT spectra of SdH measurements in \urs\  for $H \parallel c$ at the lowest temperature of $T \approx 35$\,mK for $P = 0.05$\,GPa and $T \approx 25$\,mK for $P = 1.55$\,GPa. b) Pressure dependence of the FFT frequencies. The $\beta$ peak is split as explained in the text, indicated by the dashed lines. c) Pressure dependence of the effective masses determined in a field range of 8 - 13\,T.}
\label{pdepFFT}
\end{figure}
Fig.\,\ref{pdepFFT}a presents the FFT spectra for two pressures. 
A small change in the spectra is seen for the $\beta$ branch at high pressure deep inside the AF phase. 
It is clearly split into two separate peaks with nearly the same masses. Recalling that at low pressure there is a second small peak within the $\beta$ peak, the analysis shows that the splitting increases at $P_x$ and the amplitude of the second peak increases strongly. 
The reason for the increased splitting in the AF phase may be a stronger non-linear field dependence of the SdH frequencies or a stronger warping of the corresponding Fermi surface pocket. But this needs further experimental investigations principally the angular dependence under pressure.
The pressure dependence of the FFT frequencies is shown in Fig.\,\ref{pdepFFT}b. $F_{\alpha}$ increases slightly with pressure and then has a plateau in the AF phase, in agreement with Ref. \onlinecite{Nakashima2003}. $F_{\beta}$ is, apart from the increased splitting (position of the second peak indicated by the dashed line), independent of pressure. $F_{\gamma}$ decreases with pressure and then jumps to a higher value at $P_x$. Pockets of this size are very sensitive to small changes of the band structure. 
\\These measurements indicate no significant change in the FS in between the HO phase and the AF phase. In two recent theoretical proposals the order parameter has an ordering vector $Q =$~(0~0~1) \cite{Elgazzar2009, Haule2009} and the Fermi surface in the AF and HO states are similar or the same within each model. 
All the masses decrease with pressure as seen in Fig.\,\ref{pdepFFT}c. The decrease agrees with the decrease of the $A$ coefficient of the $T^2$ behavior of $\rho -\rho_0$ with pressure \cite{Hassinger2010b}. 

To conclude, by measuring the magnetoresistance at ambient pressure for different angles between $H \parallel c$ and $H \parallel a$, we have detected the heavy branch $\eta$ and we have shown that the previously detected heavy $\beta$ branch splits into two branches when rotating the field from the $c$ to the $a$ axis. Pockets of this shape appear in band structure calculations in the AF state. Independently of calculations, under pressure for $H \parallel c$ the Fermi surface shows only minor changes between the HO state and the AF state. These are strong indications that both phases have the same unit cell doubling and the same ordering vector. 
All the detected Fermi surface pockets can account for 55\,\% of the Sommerfeld coefficient determined by specific heat measurements. 
Our accurate experimental determination of the Fermi surface is a good test for recent theoretical proposals of the HO phase when complete band structure calculations will be achieved \cite{Yamagami2000,Elgazzar2009,Haule2009,Harima2010}. Recently, the angular dependence of the SdH frequencies in the AF phase \cite{Elgazzar2009} was published \cite{Oppeneer2010} and the angular dependence of the $\alpha$ and $\beta$ branches seems in reasonable agreement with the data presented here at ambient pressure.

We acknowledge helpful discussions and calculations by L. Malone, S. Julien, A. Santander-Syro, H. Harima, G. Kotliar, H. Yamagami and P. Oppeneer. Funding came from French ANR projects DELICE, CORMAT and SINUS.

%

\end{document}